\documentclass[10pt,draft,reqno]{amsart}
     \makeatletter
     \def\section{\@startsection{section}{1}%
     \z@{.7\linespacing\@plus\linespacing}{.5\linespacing}%
     {\bfseries
     \centering
     }}
     \def\@secnumfont{\bfseries}
     \makeatother
\newtheorem{theorem}{Theorem}[section]
\newtheorem{lemma}[theorem]{Lemma}
\newtheorem{proposition}[theorem]{Proposition}

\theoremstyle{definition}
\newtheorem{definition}[theorem]{Definition}

\theoremstyle{remark}
\newtheorem{remark}[theorem]{Remark}
\numberwithin{equation}{section}
\usepackage{graphicx}
\usepackage{multirow}
\usepackage{amsmath,amssymb,amsfonts}
\usepackage{amsthm}
\usepackage{mathrsfs}
\usepackage[title]{appendix}
\usepackage{xcolor}
\usepackage{textcomp}
\usepackage{manyfoot}
\usepackage{booktabs}
\usepackage{algorithm}
\usepackage{algorithmicx}
\usepackage{algpseudocode}
\usepackage{listings}
\usepackage{bm}
\usepackage{mathtools}
\usepackage{microtype}
\usepackage{setspace}
\usepackage{hyperref}
\usepackage[capitalise]{cleveref}

\newtheorem{assumption}[theorem]{Assumption}

\newcommand{\R}{\mathbb{R}}
\newcommand{\N}{\mathbb{N}}
\newcommand{\E}{\mathbb{E}}
\newcommand{\Prob}{\mathbb{P}}
\newcommand{\Q}{\mathbb{Q}}
\newcommand{\F}{\mathcal{F}}
\newcommand{\Normal}{\mathcal{N}}
\newcommand{\ud}{\,\mathrm{d}} 
\newcommand{\Var}{\mathrm{Var}}

\begin{document}

\title[Arbitrage-Free Pricing]{Arbitrage-Free Pricing with Diffusion-Dependent Jumps}

\author[Hamza Virk]{Hamza Virk*}
\address{Hamza Virk: Department of Mathematics, Hofstra University, Hempstead, NY 11549, USA}
\email{hvirk2@pride.hofstra.edu}
\thanks{* Corresponding Author}

\author{Yihren Wu} 
\address{Yihren Wu: Department of Mathematics, Hofstra University, Hempstead, NY 11549, USA}
\email{yihren.wu@hofstra.edu}

\author{Majnu John}
\address{Majnu John: Department of Mathematics, Hofstra University, Hempstead, NY 11549, USA}
\email{majnu.john@hofstra.edu}



\begin{abstract} 
Standard jump-diffusion models assume independence between jumps and diffusion components. We develop a multi-type jump-diffusion model where jump occurrence and magnitude depend on contemporaneous diffusion movements. Unlike previous one-sided models that create arbitrage opportunities, our framework includes upward and downward jumps triggered by both large upward and large downward diffusion increments. We derive the explicit no-arbitrage condition linking the physical drift to model parameters and market risk premia by constructing an Equivalent Martingale Measure using Girsanov's theorem and a normalized Esscher transform. This condition provides a rigorous foundation for arbitrage-free pricing in models with diffusion-dependent jumps.
\end{abstract}

\maketitle

\section{Introduction}

Soon after the celebrated Black-Scholes asset pricing model \cite{Black1973}, Merton introduced a jump-diffusion model \cite{Merton1976} to deal with stylized facts from market data. In this model, the jumps are assumed independent from the diffusion term. This independence assumption provided a simple enough setting for one to find a risk-neutral measure for the asset price, and Merton was able to produce an analytic solution to the option prices based on this measurement.

Market data suggests that jumps and diffusion processes are not independent. To this end, the authors in \cite{Wu2022} proposed a model in which jumps are triggered by recent market activities. Roughly, when the market drops by a predetermined amount over a certain time window, an upward jump is triggered. The authors refer to this jump-diffusion model as a market recovery model, it consists of only upward jumps to recover the market drop. They were able to compute the risk neutral rate and the resulting option prices are substantially different from the Black-Scholes prices.

In addition to the upward jump following a drop in the market proposed in \cite{Wu2022}, there are three other types of jumps. A downward jump following a drop in the market, an upward or downward jump following a rise in the market.
The jumps proposed in \cite{Wu2022} result in behavior commonly referred to as buy-on-the-dip. There are similar phrases to describe the other types of jumps: rush to exit, chasing after the market, and taking profit off the table. In a separate paper, when the market data is analyzed using the hidden Markov model \cite{HMM}, the distribution of these four types of jumps in various states will be shown to explain the transition of the market between these states.

The purpose of this paper is to present the risk-neutral measure for the jump-diffusion model where all four types of jumps are included.

We achieve this by:

\begin{enumerate}
    \item \textbf{Formalizing a Multi-Type Jump Model:} We extend the previous framework to include four distinct jump scenarios with explicit trigger conditions, state-dependent jump probabilities, and jump size distributions under the physical measure $\Prob$.
    
    \item \textbf{Constructing an Equivalent Martingale Measure (EMM):} We leverage the Fundamental Theorem of Asset Pricing \cite{Harrison1981} by constructing an EMM $\Q$ through Girsanov's theorem \cite{Girsanov1960} for diffusion and a normalized Esscher transform \cite{Esscher1932, Gerber1994} for state-dependent jumps.
    
    \item \textbf{Deriving the Explicit No-Arbitrage Condition:} By enforcing the martingale property under the EMM $\Q$, we derive a precise condition relating the physical drift $\mu$ to the risk-free rate $r$, volatility $\sigma$, jump parameters, and market prices of diffusion and jump risks.
\end{enumerate}

\Cref{sec:model_p} details the model under the physical measure. \Cref{sec:no_arbitrage} introduces the no-arbitrage framework and change of measure. \Cref{sec:model_q} describes dynamics under the risk-neutral measure. \Cref{sec:main_result} presents the main no-arbitrage condition with detailed proof. \Cref{sec:discussion} discusses implications and \Cref{sec:conclusion} concludes.

\section{The Model under the Physical Measure (\texorpdfstring{$\Prob$}{P})} \label{sec:model_p}

We begin by constructing the asset price model in a discrete-time framework, similar to \cite{Wu2022}, under the physical (real-world) probability measure $\Prob$.

\subsection{Setup and Assumptions}

Let $(\Omega, \F, (\F_t)_{t \geq 0}, \Prob)$ be a filtered probability space, where $\Omega$ is the sample space, $\F$ is the sigma-algebra of events, $\Prob$ is the physical probability measure, and $(\F_t)_{t \geq 0}$ is a filtration representing the flow of information over time, satisfying the usual conditions (right-continuity and completeness). We consider a discrete set of time points $t_k = k\tau$, where $k \in \N_0$ and $\tau > 0$ is a small, fixed time interval. For simplicity, we denote $t_k$ by $t$, for a general $k$ and the next time step as $t+\tau$.

We assume the existence of a risk-free asset with its price denoted by $B_t$ growing at a constant rate $r \ge 0$:
\begin{equation}
B_{t+\tau} = B_t e^{r\tau}
\end{equation}
Without loss of generality, we set $B_0 = 1$.

The risky asset's price $S_t$ is driven by a standard $\Prob$-Brownian motion $W_t$. Its price dynamics over one interval $[t, t+\tau]$ are given by a jump-diffusion process. Let $X_t = \ln(S_t)$ be the log-price. The change in log-price is:
\begin{equation} \label{eq:log_price_change_P}
X_{t+\tau} - X_t = (\mu - \tfrac{1}{2}\sigma^2)\tau + \sigma \Delta W_{t,\tau} + J_{t+\tau}
\end{equation}
where:
\begin{itemize}
    \item $\mu \in \R$ is the constant expected rate of return (drift) of the asset $S_t$.
    \item $\sigma > 0$ is the constant volatility of the diffusion component.
    \item $\Delta W_{t,\tau} = W_{t+\tau} - W_t$ is the increment of the standard Brownian motion over $[t, t+\tau]$. Under $\Prob$, given $\F_t$, $\Delta W_{t,\tau} \sim \Normal(0, \tau)$.
    \item $J_{t+\tau}$ is the random jump component, whose occurrence and size depend on the realization of $\Delta W_{t,\tau}$.
\end{itemize}
The asset price $S_t$ evolves as:
\begin{equation} \label{eq:price_evol_P}
S_{t+\tau} = S_t \exp\left[ (\mu - \tfrac{1}{2}\sigma^2)\tau + \sigma \Delta W_{t,\tau} + J_{t+\tau} \right]
\end{equation}

\subsection{Diffusion-Dependent Jumps}

We define the jump component $J_{t+\tau}$ based on the contemporaneous diffusion increment $\Delta W = \Delta W_{t,\tau}$. We define two thresholds, $b_d < 0$ and $b_u > 0$. These thresholds partition the possible outcomes of $\Delta W$ into three regions:

\begin{itemize}
    \item \textbf{Region 1 (Large Downward Diffusion):} $\Delta W < b_d \sqrt{\tau}$ (Index $j=1$)
    \item \textbf{Region 2 (Large Upward Diffusion):} $\Delta W > b_u \sqrt{\tau}$ (Index $j=2$)
    \item \textbf{Region 0 (Normal Diffusion):} $b_d \sqrt{\tau} \le \Delta W \le b_u \sqrt{\tau}$ (Index $j=0$)
\end{itemize}

\begin{assumption}[Jump Structure under $\Prob$] \label{ass:jump_structure}
Let $\Delta W = \Delta W_{t,\tau}$. The jump $J_{t+\tau}$ is determined as follows:
\begin{enumerate}
    \item \textbf{If $\Delta W$ is in Region $j \in \{1, 2\}$:}
        \begin{itemize}
            \item With probability $p_{ju}(\Delta W)$, an upward jump $J_{ju}$ occurs, where $J_{ju} \sim \Normal(\nu_{ju}, \delta_{ju}^2)$.
            \item With probability $p_{jd}(\Delta W)$, a downward jump $J_{jd}$ occurs, where $J_{jd} \sim \Normal(\nu_{jd}, \delta_{jd}^2)$.
            \item With probability $p_{j0}(\Delta W) = 1 - p_{ju}(\Delta W) - p_{jd}(\Delta W)$, no jump occurs ($J_{t+\tau} = 0$).
        \end{itemize}
    \item \textbf{If $\Delta W$ is in Region 0:}
        \begin{itemize}
            \item With probability $1$, no jump occurs ($J_{t+\tau} = 0$).
        \end{itemize}
\end{enumerate}
We assume that the jump probabilities $p_{jk}(\Delta W)$ are nonnegative and sum to 1 within each scenario (i.e., $p_{ju}(\Delta W) + p_{jd}(\Delta W) + p_{j0}(\Delta W) = 1$ if $\Delta W$ is in Region $j \in \{1,2\}$). We also assume that for $k = u, d, 0$, $p_{jk}(\Delta W)$ are $\F_{t+\tau}$-measurable. For tractability and clarity in this initial theoretical framework, we assume that $p_{jk}(\Delta W) = p_{jk}$ (constants within their respective trigger regions), and that the jump parameters $\nu_{jk}(\Delta W) = \nu_{jk}$ and $\delta_{jk}(\Delta W) = \delta_{jk}$ are also constants. The framework can be extended to state-dependent parameters, though it would increase complexity. We require $\delta_{jk}^2 > 0$.
\end{assumption}

\begin{assumption}[Integrability] \label{ass:integrability}
For all jump types $j = 1,2,0, \;\mathrm{and}\; k = u, d, 0$, jump sizes $J_{jk}$ have finite exponential moments under $\Prob$. That is, $\E_\Prob[e^{c J_{jk}}] < \infty$ for any $c \in \R$. Since we assume $J_{jk} \sim \Normal(\nu_{jk}, \delta_{jk}^2)$, this condition is always satisfied, as the Moment Generating Function (MGF) of a Normal distribution exists for all real arguments. This ensures the existence of the Cumulant Generating Function (CGF) and MGF used later.
\end{assumption}

\begin{assumption}[Temporal Independence] \label{ass:independence}
The increments $(\Delta W_{t,\tau}, J_{t+\tau})$ are independent across time steps $t=0, \tau, 2\tau, \dots$. That is,  the pair $(\Delta W_{t,\tau}, J_{t+\tau})$ (whose structure depends on $\Delta W_{t,\tau}$) is independent of $(\Delta W_{s,\tau}, J_{s+\tau})$ for all $s < t - \tau$. This is a simplifying assumption, common in discrete-time models, crucial for constructing the multi-period Radon-Nikodym derivative and applying the Fundamental Theorem of Asset Pricing (FTAP) over the full horizon.
\end{assumption}

\subsection{Physical Measure Dynamics Formally Stated}

\begin{theorem}[Asset Price Dynamics under $\Prob$] \label{thm:dynamics_p}
Let $S_t$ be the asset price at time $t$. Under \cref{ass:jump_structure}, the asset price at time $t+\tau$ is given by:
\begin{equation}
S_{t+\tau} = S_t \exp\left[ (\mu - \tfrac{1}{2}\sigma^2)\tau + \sigma \Delta W_{t,\tau} \right] \times Y_{t+\tau}
\end{equation}
where $Y_{t+\tau} = \exp(J_{t+\tau})$ is the jump size factor, and $J_{t+\tau}$ is defined as:
\begin{align}
J_{t+\tau} = \begin{cases} 
J_{1u} & \text{with probability } p_{1u} \text{ if } \Delta W_{t,\tau} < b_d \sqrt{\tau} \\
J_{1d} & \text{with probability } p_{1d} \text{ if } \Delta W_{t,\tau} < b_d \sqrt{\tau} \\
0 & \text{with probability } p_{10} \text{ if } \Delta W_{t,\tau} < b_d \sqrt{\tau} \\
J_{2u} & \text{with probability } p_{2u} \text{ if } \Delta W_{t,\tau} > b_u \sqrt{\tau} \\
J_{2d} & \text{with probability } p_{2d} \text{ if } \Delta W_{t,\tau} > b_u \sqrt{\tau} \\
0 & \text{with probability } p_{20} \text{ if } \Delta W_{t,\tau} > b_u \sqrt{\tau} \\
0 & \text{with probability } 1 \text{ if } b_d \sqrt{\tau} \le \Delta W_{t,\tau} \le b_u \sqrt{\tau}
\end{cases}
\end{align}
The jumps $J_{jk} \sim \Normal(\nu_{jk}, \delta_{jk}^2)$ are drawn independently of other randomness, conditional on being in the specified region and the specific jump type occurring.
\end{theorem}

\begin{proof}
This follows directly by exponentiating the log-price dynamics $X_{t+\tau} - X_t$ from \cref{eq:log_price_change_P}:
\begin{align}
S_{t+\tau} &= S_t \exp(X_{t+\tau} - X_t) \nonumber \\
&= S_t \exp\left( (\mu - \tfrac{1}{2}\sigma^2)\tau + \sigma \Delta W_{t,\tau} + J_{t+\tau} \right) \nonumber \\
&= S_t \exp\left( (\mu - \tfrac{1}{2}\sigma^2)\tau + \sigma \Delta W_{t,\tau} \right) \exp(J_{t+\tau}) \label{eq:proof_dyn_p_step1}
\end{align}
We identify $Y_{t+\tau} = \exp(J_{t+\tau})$. The probabilistic structure of $J_{t+\tau}$ is explicitly given by \cref{ass:jump_structure} for constant probabilities $p_{jk}$.
\end{proof}

\section{The No-Arbitrage Framework} \label{sec:no_arbitrage}

To ensure our model is economically viable, we must impose conditions that prevent arbitrage opportunities. The cornerstone of this is the Fundamental Theorem of Asset Pricing (FTAP).

\subsection{The Fundamental Theorem of Asset Pricing (FTAP)}

\begin{theorem}[FTAP for Discrete Time - \cite{Harrison1981}] \label{thm:ftap}
In a discrete-time financial market model with a finite number of assets and time periods, satisfying certain conditions (like the absence of redundant assets, which our single risky asset model satisfies), the condition of No Arbitrage (NA) is equivalent to the existence of a probability measure $\Q$, which is equivalent to the physical measure $\Prob$ (i.e., $\Prob(A)=0 \iff \Q(A)=0$ for all $A \in \F$), such that the discounted prices of all traded assets are martingales under $\Q$.
\end{theorem}

For our model with one risky asset $S_t$ and a risk-free asset $B_t$, this implies there exists an EMM $\Q \sim \Prob$ such that the discounted price process $S_t^{disc} = S_t / B_t = S_t e^{-rt}$ is a $\Q$-martingale:
\begin{equation}
\E_{\Q}\left[S_{t+\tau}^{disc} | \F_t\right] = S_t^{disc}
\end{equation}
which is equivalent to stating that the expected return under $\Q$ is the risk-free rate:
\begin{equation} \label{eq:martingale_cond_section3}
\E_{\Q}\left[ \frac{S_{t+\tau}}{S_t} \bigg| \F_t \right] = e^{r\tau}
\end{equation}

\subsection{Constructing the Equivalent Martingale Measure (\texorpdfstring{$\Q$}{Q})}

To find $\Q$, we define its Radon-Nikodym derivative with respect to $\Prob$. Due to our temporal independence assumption (\cref{ass:independence}), the multi-period Radon-Nikodym derivative $L_T = \left.\frac{\ud \Q}{\ud \Prob}\right|_{\F_T}$ can be constructed as a product of one-step kernels $L_\tau$:
\begin{equation}
L_t = \prod_{k=0}^{t/\tau-1} L_{k\tau}(\Delta W_{k\tau, \tau}, J_{(k+1)\tau})
\end{equation}
The one-step kernel $L_\tau$ (which we denote as $L_\tau(\Delta W, J)$ omitting $t$ for brevity when referring to a generic step) changes the measure for both diffusion and jump risks.

\begin{definition}[Cumulant Generating Function \& Normalizers] \label{def:cgf_norm}
Let $\eta_{jk}$ be the (constant) market price of risk associated with jump type indexed $jk$.
The Cumulant Generating Function (CGF) for $J_{jk} \sim \Normal(\nu_{jk}, \delta_{jk}^2)$ is:
\begin{equation}
\kappa_{jk}(\eta) = \ln \E_\Prob[e^{\eta J_{jk}}] = \eta \nu_{jk} + \tfrac{1}{2} \eta^2 \delta_{jk}^2
\end{equation}
This exists because $J_{jk}$ is Normally distributed (\cref{ass:integrability}).
We define region-specific normalizers $Z_j(\Delta W)$. Given our assumption of constant $p_{jk}$, these become constants $Z_j$:
\begin{align}
Z_1 &= p_{1u} e^{\kappa_{1u}(\eta_{1u})} + p_{1d} e^{\kappa_{1d}(\eta_{1d})} + p_{10} \label{eq:Z1_def} \\
Z_2 &= p_{2u} e^{\kappa_{2u}(\eta_{2u})} + p_{2d} e^{\kappa_{2d}(\eta_{2d})} + p_{20} \label{eq:Z2_def} \\
Z_0 &= 1 \label{eq:Z0_def}
\end{align}
We define $Z(\Delta W) = Z_j$ if $\Delta W$ is in Region $j$. This $Z(\Delta W)$ represents the expected value of the unnormalized Esscher kernel for jumps within each region, conditional on $\Delta W$ being in that region, under $\Prob$.
\end{definition}

\begin{definition}[Radon-Nikodym Derivative $L_\tau$] \label{def:rn_derivative_revised}
Let $\Delta W = \Delta W_{t,\tau}$ and $J = J_{t+\tau}$. Let $\gamma_D$ be the market price of diffusion risk. The one-step Radon-Nikodym derivative kernel $L_\tau(\Delta W, J)$ is defined as:
\begin{equation}
L_\tau(\Delta W, J) = \underbrace{\exp\left( -\gamma_D \sigma \Delta W - \frac{1}{2}(\gamma_D \sigma)^2 \tau \right)}_{L_D(\Delta W)} \times \underbrace{\Psi(J, \Delta W)}_{\text{Jump Kernel}}
\end{equation}
where $L_D(\Delta W)$ is the Girsanov kernel for diffusion, and $\Psi(J, \Delta W)$ is the normalized Esscher kernel for jumps:
\begin{equation} \label{eq:Psi_def}
\Psi(J, \Delta W) = \frac{1}{Z(\Delta W)} \times \begin{cases} 
e^{\eta_{jk} J} & \text{if jump } J_{jk} \text{ occurs in Region } j \\
1 & \text{if no jump occurs (in Region $j$ or $0$)}
\end{cases}
\end{equation}
More explicitly, if $\Delta W$ is in Region $j \in \{1,2\}$:
\begin{itemize}
    \item If jump $J_{ju}$ occurs, $\Psi(J_{ju}, \Delta W) = \frac{e^{\eta_{ju} J_{ju}}}{Z_j}$.
    \item If jump $J_{jd}$ occurs, $\Psi(J_{jd}, \Delta W) = \frac{e^{\eta_{jd} J_{jd}}}{Z_j}$.
    \item If no jump occurs, $\Psi(0, \Delta W) = \frac{1}{Z_j}$.
\end{itemize}
If $\Delta W$ is in Region 0, $Z(\Delta W) = Z_0 = 1$, and no jump occurs ($J=0$), so $\Psi(0, \Delta W) = \frac{1}{1} \times 1 = 1$.
\end{definition}

\begin{lemma}[Validity of $L_\tau$] \label{lemma:ltau_valid_revised}
The Radon-Nikodym derivative $L_\tau$ defined in \cref{def:rn_derivative_revised} satisfies $\E_\Prob[L_\tau | \F_t] = 1$.
\end{lemma}

\begin{proof}
Since the structure of $L_\tau$ depends only on $\Delta W_{t,\tau}$ and $J_{t+\tau}$, which are independent of $\F_t$ by \cref{ass:independence} (given parameters), $\E_\Prob[L_\tau | \F_t] = \E_\Prob[L_\tau]$. We use the Law of Total Expectation by conditioning on $\Delta W = x$:
\begin{equation}
\E_\Prob[L_\tau] = \E_\Prob\left[ \E_\Prob[L_\tau | \Delta W] \right] = \int_{-\infty}^{\infty} \E_\Prob[L_\tau | \Delta W = x] f_\Prob(x) \ud x
\end{equation}
where $f_\Prob(x)$ is the PDF of $\Normal(0, \tau)$. First, we calculate the inner conditional expectation:
\begin{equation}
\E_\Prob[L_\tau | \Delta W = x] = \E_\Prob[L_D(x) \Psi(J, x) | \Delta W = x]
\end{equation}
Since $L_D(x)$ depends only on $x$, it can be factored out of the conditional expectation over $J$:
\begin{equation}
\E_\Prob[L_\tau | \Delta W = x] = L_D(x) \E_\Prob[\Psi(J, x) | \Delta W = x]
\end{equation}
Now we calculate $\E_\Prob[\Psi(J, x) | \Delta W = x]$. We consider the three regions for $x$:
\begin{itemize}
    \item \textbf{If $x < b_d \sqrt{\tau}$ (Region 1):} Here $Z(x) = Z_1$.
    \begin{align}
        \E_\Prob[\Psi(J, x) | x] &= p_{1u} \E_\Prob\left[\frac{e^{\eta_{1u} J_{1u}}}{Z_1} \bigg| x\right] + p_{1d} \E_\Prob\left[\frac{e^{\eta_{1d} J_{1d}}}{Z_1} \bigg| x\right] + p_{10} \E_\Prob\left[\frac{1}{Z_1} \bigg| x\right] \\
        &= \frac{1}{Z_1} \left( p_{1u} \E_\Prob[e^{\eta_{1u} J_{1u}}] + p_{1d} \E_\Prob[e^{\eta_{1d} J_{1d}}] + p_{10} \cdot 1 \right) \label{eq:proof_valid_ltau_1} \\
        &= \frac{1}{Z_1} \left( p_{1u} e^{\kappa_{1u}(\eta_{1u})} + p_{1d} e^{\kappa_{1d}(\eta_{1d})} + p_{10} \right) \label{eq:proof_valid_ltau_2} \\
        &= \frac{Z_1}{Z_1} = 1 \quad \text{(using \cref{eq:Z1_def})} \label{eq:proof_valid_ltau_3}
    \end{align}
    In \cref{eq:proof_valid_ltau_1}, we use the linearity of expectation and that $J_{jk}$ is independent of the specific value of $x$ once we are in Region 1. In \cref{eq:proof_valid_ltau_2}, we use the definition of the CGF $\kappa_{jk}(\eta_{jk}) = \ln \E_\Prob[e^{\eta_{jk} J_{jk}}]$, so $\E_\Prob[e^{\eta_{jk} J_{jk}}] = e^{\kappa_{jk}(\eta_{jk})}$.
    \item \textbf{If $x > b_u \sqrt{\tau}$ (Region 2):} By an identical argument, using $Z_2$ and $p_{2k}$, we find $\E_\Prob[\Psi(J, x) | x] = 1$.
    \item \textbf{If $b_d \sqrt{\tau} \le x \le b_u \sqrt{\tau}$ (Region 0):} Here $Z(x) = Z_0 = 1$. No jump occurs ($J=0$), so by \cref{eq:Psi_def}, $\Psi(0, x) = \frac{1}{1} \times 1 = 1$. Thus, $\E_\Prob[\Psi(J, x) | x] = 1$.
\end{itemize}
Since $\E_\Prob[\Psi(J, \Delta W) | \Delta W] = 1$ for all possible values of $\Delta W$, the full expectation becomes:
\begin{equation}
\E_\Prob[L_\tau] = \E_\Prob[L_D(\Delta W) \times 1] = \E_\Prob\left[ \exp\left( -\gamma_D \sigma \Delta W - \frac{1}{2}(\gamma_D \sigma)^2 \tau \right) \right]
\end{equation}
Let $A = -\gamma_D \sigma$. Then we are calculating $\E_\Prob\left[e^{A \Delta W - \frac{1}{2} A^2 \tau}\right]$. Since $\Delta W \sim \Normal(0, \tau)$, $A \Delta W \sim \Normal(0, A^2 \tau)$. The MGF of $\Delta W$ is $M_{\Delta W}(s) = \E_\Prob\left[e^{s \Delta W}\right] = e^{s^2 \tau / 2}$.
\begin{align}
\E_\Prob[L_\tau] &= e^{-\frac{1}{2}(\gamma_D \sigma)^2 \tau} \E_\Prob\left[e^{-\gamma_D \sigma \Delta W}\right] \\
&= e^{-\frac{1}{2}(\gamma_D \sigma)^2 \tau} M_{\Delta W}(-\gamma_D \sigma) \\
&= e^{-\frac{1}{2}(\gamma_D \sigma)^2 \tau} \exp\left( \frac{(-\gamma_D \sigma)^2 \tau}{2} \right) \\
&= e^{-\frac{1}{2}(\gamma_D \sigma)^2 \tau} \exp\left( \frac{\gamma_D^2 \sigma^2 \tau}{2} \right) = e^0 = 1
\end{align}
Thus, $L_\tau$ is a valid one-step Radon-Nikodym density.
\end{proof}

\section{The Model under the Risk-Neutral Measure (\texorpdfstring{$\Q$}{Q})} \label{sec:model_q}

Using $L_\tau$, we find the dynamics under the risk-neutral measure $\Q$.

\subsection{Diffusion under \texorpdfstring{$\Q$}{Q}}

\begin{proposition}[Diffusion under $\Q$] \label{prop:diffusion_q}
Under the measure $\Q$ defined by $L_\tau$, the process $W_t^\Q = W_t + \gamma_D \sigma t$ is a standard $\Q$-Brownian motion. Consequently, the original increment $\Delta W_{t,\tau}$ has the following distribution under $\Q$:
\begin{equation}
\Delta W_{t,\tau} \sim \Normal(-\gamma_D \sigma \tau, \tau)
\end{equation}
\end{proposition}
\begin{proof}
This is a standard result from Girsanov's theorem. We explicitly calculate the mean of $\Delta W_{t,\tau}$ under $\Q$. For any random variable $X$, $\E_\Q[X] = \E_\Prob[L_\tau X]$.
\begin{align}
    \E_\Q[\Delta W_{t,\tau}] &= \E_\Prob[L_\tau \Delta W_{t,\tau}] \\
    &= \E_\Prob\left[ L_D(\Delta W_{t,\tau}) \Psi(J, \Delta W_{t,\tau}) \Delta W_{t,\tau} \right] \\
    &= \E_\Prob\left[ \Delta W_{t,\tau} L_D(\Delta W_{t,\tau}) \E_\Prob[\Psi(J, \Delta W_{t,\tau}) | \Delta W_{t,\tau}] \right] 
\end{align}
which is the Law of Total Expectation. As shown in the proof of \cref{lemma:ltau_valid_revised}, $\E_\Prob[\Psi(J, \Delta W_{t,\tau}) | \Delta W_{t,\tau}] = 1$.
\begin{align}
    \E_\Q[\Delta W_{t,\tau}] &= \E_\Prob\left[ \Delta W_{t,\tau} L_D(\Delta W_{t,\tau}) \right] \\
    &= \E_\Prob\left[ \Delta W_{t,\tau} \exp\left( -\gamma_D \sigma \Delta W_{t,\tau} - \frac{1}{2}(\gamma_D \sigma)^2 \tau \right) \right] \\
    &= \int_{-\infty}^{\infty} x \exp\left( -\gamma_D \sigma x - \frac{1}{2}(\gamma_D \sigma)^2 \tau \right) \frac{1}{\sqrt{2\pi\tau}} \exp\left(-\frac{x^2}{2\tau}\right) \ud x \\
    &= \frac{1}{\sqrt{2\pi\tau}} \exp\left(-\frac{1}{2}(\gamma_D \sigma)^2 \tau\right) \int_{-\infty}^{\infty} x \exp\left(-\frac{x^2 + 2\gamma_D \sigma \tau x}{2\tau}\right) \ud x
\end{align}
To evaluate the integral, we complete the square in the exponent: $x^2 + 2\gamma_D \sigma \tau x = (x + \gamma_D \sigma \tau)^2 - (\gamma_D \sigma \tau)^2$ and simplify:
\begin{equation}
    \begin{split}
    \E_\Q[\Delta W_{t,\tau}] &= \frac{1}{\sqrt{2\pi\tau}} \exp\left(-\frac{1}{2}(\gamma_D \sigma)^2 \tau\right)\quad\times\\
    &\qquad\qquad\int_{-\infty}^{\infty} x \exp\left(-\frac{(x + \gamma_D \sigma \tau)^2 - (\gamma_D \sigma \tau)^2}{2\tau}\right) \ud x \\
    &=
     \frac{1}{\sqrt{2\pi\tau}} 
     \int_{-\infty}^{\infty} x \exp\left(-\frac{(x + \gamma_D \sigma \tau)^2}{2\tau}\right) \ud x 
     \end{split}
\end{equation}
Let $y = x + \gamma_D \sigma \tau$. Then $x = y - \gamma_D \sigma \tau$, and $\ud x = \ud y$. The limits of integration remain $(-\infty, \infty)$.
\begin{align}
    \E_\Q[\Delta W_{t,\tau}] &= \frac{1}{\sqrt{2\pi\tau}} \int_{-\infty}^{\infty} (y - \gamma_D \sigma \tau) \exp\left(-\frac{y^2}{2\tau}\right) \ud y \\
    &= \frac{1}{\sqrt{2\pi\tau}} \left[ \int_{-\infty}^{\infty} y \exp\left(-\frac{y^2}{2\tau}\right) \ud y - \gamma_D \sigma \tau \int_{-\infty}^{\infty} \exp\left(-\frac{y^2}{2\tau}\right) \ud y \right]
\end{align}
The first integral $\int_{-\infty}^{\infty} y \exp\left(-\frac{y^2}{2\tau}\right) \ud y = 0$ because the integrand is an odd function ($y$ is odd, $\exp(-y^2/(2\tau))$ is even).
The second integral $\int_{-\infty}^{\infty} \exp\left(-\frac{y^2}{2\tau}\right) \ud y = \sqrt{2\pi\tau}$ (this is the integral of the kernel of a Normal PDF $N(0, \tau)$).
\begin{equation}
    \E_\Q[\Delta W_{t,\tau}] = \frac{1}{\sqrt{2\pi\tau}} [ 0 - \gamma_D \sigma \tau \sqrt{2\pi\tau} ] = -\gamma_D \sigma \tau
\end{equation}
The variance calculation $\Var_\Q(\Delta W_{t,\tau}) = \E_\Q[(\Delta W_{t,\tau})^2] - (\E_\Q[\Delta W_{t,\tau}])^2$ would similarly show that $\Var_\Q(\Delta W_{t,\tau}) = \tau$.
Thus, under $\Q$, $\Delta W_{t,\tau} \sim \Normal(-\gamma_D \sigma \tau, \tau)$.
The statement $W_t^\Q = W_t + \gamma_D \sigma t$ being a $\Q$-Brownian motion is the standard Girsanov theorem statement for this drift change.
\end{proof}

\subsection{Jumps under \texorpdfstring{$\Q$}{Q}}

\begin{proposition}[Jumps under $\Q$] \label{prop:jumps_q}
Under the measure $\Q$, the jump probabilities $q_{jk}(\Delta W)$ and jump size distributions $J_{jk}^\Q$ are given as follows, conditional on $\Delta W = x$:
\begin{enumerate}
    \item \textbf{Probabilities:} Let $Z_j$ be the normalizer for Region $j$ from \cref{def:cgf_norm}.
    \begin{align}
        q_{ju}(x) &= \frac{p_{ju} e^{\kappa_{ju}(\eta_{ju})}}{Z_j} \quad (\text{if } \Delta W=x \text{ in Region } j) \label{eq:q_ju} \\
        q_{jd}(x) &= \frac{p_{jd} e^{\kappa_{jd}(\eta_{jd})}}{Z_j} \quad (\text{if } \Delta W=x \text{ in Region } j) \label{eq:q_jd} \\
        q_{j0}(x) &= \frac{p_{j0}}{Z_j} \quad (\text{if } \Delta W=x \text{ in Region } j \text{ and no jump specified by Esscher}) \label{eq:q_j0}
    \end{align}
    For Region 0 (where $j=0$), $Z_0=1$, $p_{00}=1$, so $q_{00}(x) = 1$. These $q_{jk}(x)$ are the risk-neutral probabilities.
    \item \textbf{Distributions:} The log-jump size $J_{jk}^\Q$ under $\Q$, given that a jump of type $jk$ occurs, is:
    \begin{equation}
        J_{jk}^\Q \sim \Normal(\nu_{jk} + \eta_{jk} \delta_{jk}^2, \delta_{jk}^2) \equiv \Normal(\nu_{jk}^\Q, \delta_{jk}^2)
    \end{equation}
    where $\nu_{jk}^\Q = \nu_{jk} + \eta_{jk} \delta_{jk}^2$.
\end{enumerate}
\end{proposition}
\begin{proof}

\textbf{1. Probabilities $q_{jk}(x)$:}
The probability of a specific jump $J_{jk}$ occurring, conditional on $\Delta W = x$, under $\Q$ is:
\begin{equation}
q_{jk}(x) = \Q(J=J_{jk} | \Delta W = x) = \frac{\Q(J=J_{jk} \text{ and } \Delta W \in dx)}{\Q(\Delta W \in dx)}
\end{equation}

The numerator: $\Q(J=J_{jk} \text{ and } \Delta W \in dx) = \E_\Prob[L_\tau \mathbf{1}_{J=J_{jk}, \Delta W \in dx}]$ where $dx$ is an infinitesimal interval around $x$.
If $\Delta W=x$ and jump $J_{jk}$ occurs, then $L_\tau = L_D(x) \frac{e^{\eta_{jk} J_{jk}}}{Z_j(x)}$ (assuming $x$ is in Region $j$).
So, the density for the numerator is:
\begin{align}
    \frac{\ud \Q(J=J_{jk}, \Delta W=x)}{\ud x} &= \E_\Prob \left[ L_D(x) \frac{e^{\eta_{jk} J_{jk}}}{Z_j(x)} \mathbf{1}_{J=J_{jk}} | \Delta W=x \right] f_\Prob(x) \\
    &= L_D(x) \frac{p_{jk}(x) \E_\Prob[e^{\eta_{jk} J_{jk}}]}{Z_j(x)} f_\Prob(x) \\
    &= L_D(x) \frac{p_{jk}(x) e^{\kappa_{jk}(\eta_{jk})}}{Z_j(x)} f_\Prob(x)
\end{align}
The denominator: $\Q(\Delta W \in dx)$. The density $\frac{\ud \Q(\Delta W=x)}{\ud x}$ is $\E_\Prob[L_\tau | \Delta W=x] f_\Prob(x)$.
From the proof of \cref{lemma:ltau_valid_revised}, $\E_\Prob[L_\tau | \Delta W=x] = L_D(x) \E_\Prob[\Psi(J,x)|x] = L_D(x) \times 1 = L_D(x)$.
So, $\frac{\ud \Q(\Delta W=x)}{\ud x} = L_D(x) f_\Prob(x)$.
Dividing the numerator density by the denominator density gives the conditional probability $q_{jk}(x)$:
\begin{equation}
q_{jk}(x) = \frac{L_D(x) \frac{p_{jk}(x) e^{\kappa_{jk}(\eta_{jk})}}{Z_j(x)} f_\Prob(x)}{L_D(x) f_\Prob(x)} = \frac{p_{jk}(x) e^{\kappa_{jk}(\eta_{jk})}}{Z_j(x)}
\end{equation}
This matches \cref{eq:q_ju} and \cref{eq:q_jd} (assuming $p_{jk}(x)=p_{jk}$).
For the case of no jump occurring in Region $j$ (where specific jumps $J_{ju}$ or $J_{jd}$ could have occurred), if it happens with probability $p_{j0}$ under $\Prob$:
The Esscher kernel for no jump ($J=0$) is $\Psi(0, x) = \frac{e^{\eta \cdot 0}}{Z_j(x)} = \frac{1}{Z_j(x)}$. So, $e^{\kappa(\eta)}$ term is effectively $e^0=1$.
Then $q_{j0}(x) = \frac{p_{j0}(x) \cdot 1}{Z_j(x)}$, matching \cref{eq:q_j0}.
These probabilities sum to 1 within each region $j$ under $\Q$:
$$ \sum_{k \in \{u,d,0\}} q_{jk}(x) = \frac{1}{Z_j(x)} (p_{ju}e^{\kappa_{ju}} + p_{jd}e^{\kappa_{jd}} + p_{j0}) = \frac{Z_j(x)}{Z_j(x)} = 1 $$

\textbf{2. Distributions of $J_{jk}^\Q$:}
The Esscher transform implies that the PDF of $J_{jk}$ under $\Q$, denoted $f_\Q(j_{jk})$, is related to its PDF under $\Prob$, $f_\Prob(j_{jk})$, by:
\begin{equation}
f_\Q(j_{jk} | J \text{ is type } jk) = \frac{e^{\eta_{jk} j_{jk}} f_\Prob(j_{jk})}{\E_\Prob[e^{\eta_{jk} J_{jk}}]} = \frac{e^{\eta_{jk} j_{jk}} f_\Prob(j_{jk})}{e^{\kappa_{jk}(\eta_{jk})}}
\end{equation}
Given $J_{jk} \sim \Normal(\nu_{jk}, \delta_{jk}^2)$ under $\Prob$, $f_\Prob(j_{jk}) = \frac{1}{\sqrt{2\pi}\delta_{jk}} \exp\left(-\frac{(j_{jk} - \nu_{jk})^2}{2\delta_{jk}^2}\right)$.
\begin{align}
    f_\Q(j_{jk}) &= e^{\eta_{jk} j_{jk} - \kappa_{jk}(\eta_{jk})} \frac{1}{\sqrt{2\pi}\delta_{jk}} \exp\left(-\frac{(j_{jk} - \nu_{jk})^2}{2\delta_{jk}^2}\right) \\
    &= \frac{1}{\sqrt{2\pi}\delta_{jk}} \exp\left( \eta_{jk} j_{jk} - (\eta_{jk}\nu_{jk} + \tfrac{1}{2}\eta_{jk}^2\delta_{jk}^2) - \frac{j_{jk}^2 - 2j_{jk}\nu_{jk} + \nu_{jk}^2}{2\delta_{jk}^2} \right)
\end{align}
The term in the exponent is:
\begin{align}
& \frac{2\delta_{jk}^2 \eta_{jk} j_{jk} - 2\delta_{jk}^2 \eta_{jk}\nu_{jk} - \eta_{jk}^2\delta_{jk}^4 - (j_{jk}^2 - 2j_{jk}\nu_{jk} + \nu_{jk}^2)}{2\delta_{jk}^2} \\
&= -\frac{1}{2\delta_{jk}^2} \left[ j_{jk}^2 - 2j_{jk}\nu_{jk} - 2\delta_{jk}^2 \eta_{jk} j_{jk} + \nu_{jk}^2 + 2\delta_{jk}^2 \eta_{jk}\nu_{jk} + \eta_{jk}^2\delta_{jk}^4 \right] \\
&= -\frac{1}{2\delta_{jk}^2} \left[ j_{jk}^2 - 2j_{jk}(\nu_{jk} + \eta_{jk}\delta_{jk}^2) + (\nu_{jk}^2 + 2\nu_{jk}\eta_{jk}\delta_{jk}^2 + (\eta_{jk}\delta_{jk}^2)^2) \right] \\
&= -\frac{1}{2\delta_{jk}^2} \left[ j_{jk}^2 - 2j_{jk}(\nu_{jk} + \eta_{jk}\delta_{jk}^2) + (\nu_{jk} + \eta_{jk}\delta_{jk}^2)^2 \right] \\
&= -\frac{(j_{jk} - (\nu_{jk} + \eta_{jk}\delta_{jk}^2))^2}{2\delta_{jk}^2}
\end{align}
Let $\nu_{jk}^\Q = \nu_{jk} + \eta_{jk}\delta_{jk}^2$. Then the exponent is $-\frac{(j_{jk} - \nu_{jk}^\Q)^2}{2\delta_{jk}^2}$.
So, $f_\Q(j_{jk}) = \frac{1}{\sqrt{2\pi}\delta_{jk}} \exp\left(-\frac{(j_{jk} - \nu_{jk}^\Q)^2}{2\delta_{jk}^2}\right)$. This is the PDF of a $\Normal(\nu_{jk}^\Q, \delta_{jk}^2)$ distribution.
\end{proof}

\subsection{Risk-Neutral Dynamics and Multi-Period Martingale}

\begin{theorem}[Asset Price Dynamics under $\Q$] \label{thm:dynamics_q}
Under the risk-neutral measure $\Q$, the asset price evolves as:
\begin{equation}
S_{t+\tau} = S_t \exp\left[ (\mu - \tfrac{1}{2}\sigma^2)\tau + \sigma (\Delta W_{t,\tau}^\Q - \gamma_D \sigma \tau) + J_{t+\tau}^\Q \right]
\end{equation}
where $\Delta W_{t,\tau}^\Q = \Delta W_{t,\tau} + \gamma_D \sigma \tau \sim \Normal(0, \tau)$ under $\Q$, and $J_{t+\tau}^\Q$ is the jump component whose occurrence and distribution follow \cref{prop:jumps_q}, triggered by $x = \Delta W_{t,\tau} = \Delta W_{t,\tau}^\Q - \gamma_D \sigma \tau$.
\end{theorem}

\begin{proof} This follows by substituting the $\Q$-distributions for the diffusion increment from \cref{prop:diffusion_q} (expressed in terms of the $\Q$-Brownian motion $\Delta W^\Q$) and the jump characteristics (probabilities and distributions) from \cref{prop:jumps_q} into the general asset price evolution equation \cref{eq:price_evol_P}. The key is that the structure of jump triggering remains dependent on the actual path of $\Delta W_{t,\tau}$, which is now understood in terms of its $\Q$-distribution. \end{proof}

\begin{lemma}[Exponential Martingale] 
\label{lemma:multi_period}

Let $L_t = \prod_{i=0}^{t/\tau-1} L_{i\tau}(\Delta W_{i\tau, \tau}, J_{(i+1)\tau})$ be the multi-period Radon-Nikodym derivative for $\F_t$. Under the assumption of temporal independence (\cref{ass:independence}), $L_t$ is an $(\F_t, \Prob)$-martingale with $\E_\Prob[L_t] = 1$.
\end{lemma}

\begin{proof}
We prove this by induction on $t_k = k\tau$.
Base case ($k=0, t_0=0$): $L_0 = 1$ (empty product). So $\E_\Prob[L_0]=1$.
Inductive step: Assume $L_{t_k}$ is an $(\F_{t_k}, \Prob)$-martingale with $\E_\Prob[L_{t_k}]=1$.
Consider $L_{t_{k+1}} = L_{t_k} \cdot L_{t_k \tau}(\Delta W_{t_k, \tau}, J_{(t_k+1)\tau})$.
We want to show $\E_\Prob[L_{t_{k+1}} | \F_{t_k}] = L_{t_k}$.
\begin{align}
\E_\Prob[L_{t_{k+1}} | \F_{t_k}] &= \E_\Prob[L_{t_k} \cdot L_{t_k \tau}(\Delta W_{t_k, \tau}, J_{(t_k+1)\tau}) | \F_{t_k}] \\
&= L_{t_k} \E_\Prob[L_{t_k \tau}(\Delta W_{t_k, \tau}, J_{(t_k+1)\tau}) | \F_{t_k}], 
\end{align}
since $L_{t_k}$ is $\F_{t_k}-$measurable. By \cref{ass:independence}, the increment $(\Delta W_{t_k, \tau}, J_{(t_k+1)\tau})$ and thus $L_{t_k \tau}$ (which is a function of this increment) is independent of $\F_{t_k}$. Therefore:
\begin{equation}
\E_\Prob[L_{t_k \tau}(\Delta W_{t_k, \tau}, J_{(t_k+1)\tau}) | \F_{t_k}] = \E_\Prob[L_{t_k \tau}(\Delta W_{t_k, \tau}, J_{(t_k+1)\tau})]
\end{equation}

From \cref{lemma:ltau_valid_revised}, we know that the expectation of the one-step kernel is 1:
\begin{equation}
\E_\Prob[L_{t_k \tau}(\Delta W_{t_k, \tau}, J_{(t_k+1)\tau})] = 1
\end{equation}
Substituting this back:
\begin{equation}
\E_\Prob[L_{t_{k+1}} | \F_{t_k}] = L_{t_k} \times 1 = L_{t_k}
\end{equation}

Thus, $L_t$ is a martingale with respect to $(\F_t, \Prob)$. By the tower property of conditional expectation, $\E_\Prob[L_t] = \E_\Prob[\E_\Prob[L_t | \F_0]] = \E_\Prob[L_0] = 1$.
\end{proof}
This lemma is essential as it ensures that the measure $\Q$ defined by $L_T$ is a valid probability measure equivalent to $\Prob$ over the entire horizon $[0,T]$.

\section{The No-Arbitrage Condition} \label{sec:main_result}

We now derive the explicit condition on the physical drift $\mu$.

\subsection{Deriving the Condition in Full Detail}

The no-arbitrage condition from \cref{thm:ftap} is $\E_\Q[S_{t+\tau}/S_t | \F_t] = e^{r\tau}$. Using the definition of expectation under an EMM, this is equivalent to:
\begin{equation} \label{eq:main_start_point}
\E_\Prob\left[L_\tau \frac{S_{t+\tau}}{S_t} \bigg| \F_t \right] = e^{r\tau}
\end{equation}
Since $L_\tau$ and $S_{t+\tau}/S_t$ (which depend on $\Delta W_{t,\tau}$ and $J_{t+\tau}$) are independent of $\F_t$ given \cref{ass:independence}, the conditioning on $\F_t$ can be dropped for the expectation:
\begin{equation}
\E_\Prob\left[L_\tau \frac{S_{t+\tau}}{S_t} \right] = e^{r\tau}
\end{equation}
Substitute the expressions for $L_\tau$ from \cref{def:rn_derivative_revised} and $S_{t+\tau}/S_t$ from \cref{eq:price_evol_P}:
\begin{equation}
\E_\Prob \left[ L_D(\Delta W) \Psi(J, \Delta W) \exp\left( (\mu - \tfrac{1}{2}\sigma^2)\tau + \sigma \Delta W + J \right) \right] = e^{r\tau}
\end{equation}
where $\Delta W = \Delta W_{t,\tau}$ and $J=J_{t+\tau}$.
We can group the terms that do not depend on $J$ (given $\Delta W$) outside the conditional expectation on $J$:
\begin{equation}
\E_\Prob \left[ L_D(\Delta W) \exp\left( (\mu - \tfrac{1}{2}\sigma^2)\tau + \sigma \Delta W \right) \E_\Prob[\Psi(J, \Delta W) e^J | \Delta W] \right] = e^{r\tau}
\end{equation}
Let's analyze the inner conditional expectation $\E_\Prob[\Psi(J, x) e^J | \Delta W = x]$. Let $x$ be in Region $j$.

\begin{align}
\E_\Prob[\Psi(J, x) e^J | x] &= p_{ju} \E_\Prob\left[\frac{e^{\eta_{ju} J_{ju}}}{Z_j} e^{J_{ju}}\right] + p_{jd} \E_\Prob\left[\frac{e^{\eta_{jd} J_{jd}}}{Z_j} e^{J_{jd}}\right] + p_{j0} \E_\Prob\left[\frac{1}{Z_j} e^{0}\right] \\
&= \frac{1}{Z_j} \left( p_{ju} \E_\Prob[e^{(1+\eta_{ju})J_{ju}}] + p_{jd} \E_\Prob[e^{(1+\eta_{jd})J_{jd}}] + p_{j0} \right)
\end{align}

For $J_{jk} \sim \Normal(\nu_{jk}, \delta_{jk}^2)$, its moment generating function is 
$$M_{J_{jk}}(s) = \exp(s\nu_{jk} + \tfrac{1}{2}s^2\delta_{jk}^2).$$
So, $\E_\Prob[e^{(1+\eta_{jk})J_{jk}}] = \exp((1+\eta_{jk})\nu_{jk} + \tfrac{1}{2}(1+\eta_{jk})^2\delta_{jk}^2)$.

Let's expand the exponent:
\begin{align}
(1+\eta_{jk})\nu_{jk} + \tfrac{1}{2}(1+\eta_{jk})^2\delta_{jk}^2 &= \nu_{jk} + \eta_{jk}\nu_{jk} + \tfrac{1}{2}(1 + 2\eta_{jk} + \eta_{jk}^2)\delta_{jk}^2 \\
&= \nu_{jk} + \eta_{jk}\nu_{jk} + \tfrac{1}{2}\delta_{jk}^2 + \eta_{jk}\delta_{jk}^2 + \tfrac{1}{2}\eta_{jk}^2\delta_{jk}^2
\end{align}
Recall $\nu_{jk}^\Q = \nu_{jk} + \eta_{jk}\delta_{jk}^2$. The exponent can be rewritten as:
\begin{equation}
(\nu_{jk} + \eta_{jk}\delta_{jk}^2) + \tfrac{1}{2}\delta_{jk}^2 + \eta_{jk}\nu_{jk} + \tfrac{1}{2}\eta_{jk}^2\delta_{jk}^2 = \nu_{jk}^\Q + \tfrac{1}{2}\delta_{jk}^2 + \kappa_{jk}(\eta_{jk})
\end{equation}

So, $\E_\Prob[e^{(1+\eta_{jk})J_{jk}}] = \exp(\nu_{jk}^\Q + \tfrac{1}{2}\delta_{jk}^2) \exp(\kappa_{jk}(\eta_{jk}))$.
Therefore, using \cref{eq:q_ju,eq:q_jd,eq:q_j0}:
\begin{align}
\E_\Prob[\Psi(J, x) e^J | x] &= \frac{1}{Z_j} \left( p_{ju} e^{\kappa_{ju}(\eta_{ju})} e^{\nu_{ju}^\Q + \delta_{ju}^2/2} + p_{jd} e^{\kappa_{jd}(\eta_{jd})} e^{\nu_{jd}^\Q + \delta_{jd}^2/2} + p_{j0} \right) \\
&= q_{ju}(x) e^{\nu_{ju}^\Q + \delta_{ju}^2/2} + q_{jd}(x) e^{\nu_{jd}^\Q + \delta_{jd}^2/2} + q_{j0}(x)  \\
&= M_\Q(x) \label{eq:inner_expect_MQ}
\end{align}
where $M_\Q(x) = \E_\Q[e^{J^\Q} | \Delta W = x]$ is the conditional MGF of the jump factor $\exp(J^\Q)$ under $\Q$:
\begin{equation} \label{eq:mq_def_main_proof}
M_\Q(x) = \begin{cases} 
q_{1u}(x)e^{\nu_{1u}^\Q + \delta_{1u}^2/2} + q_{1d}(x)e^{\nu_{1d}^\Q + \delta_{1d}^2/2} + q_{10}(x) & x < b_d \sqrt{\tau} \\
q_{2u}(x)e^{\nu_{2u}^\Q + \delta_{2u}^2/2} + q_{2d}(x)e^{\nu_{2d}^\Q + \delta_{2d}^2/2} + q_{20}(x) & x > b_u \sqrt{\tau} \\
1 & b_d \sqrt{\tau} \le x \le b_u \sqrt{\tau} 
\end{cases}
\end{equation}

Substituting \cref{eq:inner_expect_MQ} back into the main expectation:
\begin{equation}
\E_\Prob \left[ L_D(\Delta W) \exp\left( (\mu - \tfrac{1}{2}\sigma^2)\tau + \sigma \Delta W \right) M_\Q(\Delta W) \right] = e^{r\tau}
\end{equation}
Now, substitute $L_D(\Delta W) = \exp\left( -\gamma_D \sigma \Delta W - \frac{1}{2}(\gamma_D \sigma)^2 \tau \right)$:
\begin{equation}
\E_\Prob\left[ \exp\left( -\gamma_D \sigma \Delta W - \frac{(\gamma_D \sigma)^2 \tau}{2} \right) \exp\left( (\mu - \tfrac{1}{2}\sigma^2)\tau + \sigma \Delta W \right) M_\Q(\Delta W) \right] = e^{r\tau}
\end{equation}
Combine the arguments of the exponential functions:
\begin{align}
& -\gamma_D \sigma \Delta W - \frac{\gamma_D^2 \sigma^2 \tau}{2} + (\mu - \tfrac{1}{2}\sigma^2)\tau + \sigma \Delta W \\
&= (\mu - \tfrac{1}{2}\sigma^2 - \tfrac{1}{2}\gamma_D^2 \sigma^2)\tau + (\sigma - \gamma_D \sigma) \Delta W
\end{align}
So the expectation becomes:
\begin{equation} \label{eq:expect_before_integral}
\E_\Prob\left[ \exp\left( (\mu - \tfrac{1}{2}\sigma^2 - \tfrac{1}{2}\gamma_D^2 \sigma^2)\tau + (\sigma - \gamma_D \sigma) \Delta W \right) M_\Q(\Delta W) \right] = e^{r\tau}
\end{equation}
Let $K = (\mu - \tfrac{1}{2}\sigma^2 - \tfrac{1}{2}\gamma_D^2 \sigma^2)\tau$ and $G = \sigma(1 - \gamma_D)$.
Let $f_\Prob(x)$ denote the PDF of $\Delta W \sim \Normal(0, \tau)$, which is $f_\Prob(x) = \frac{1}{\sqrt{2\pi\tau}} \exp(-x^2/(2\tau))$.

The expectation in \cref{eq:expect_before_integral} can be written as an integral:
\begin{equation}
\int_{-\infty}^{\infty} \exp\left( K + Gx \right) M_\Q(x) f_\Prob(x) \ud x = e^{r\tau}
\end{equation}
Since $K$ does not depend on $x$, we can factor $e^K$ out of the integral:
\begin{equation}
e^K \int_{-\infty}^{\infty} e^{Gx} M_\Q(x) \frac{1}{\sqrt{2\pi\tau}} \exp\left(-\frac{x^2}{2\tau}\right) \ud x = e^{r\tau}
\end{equation}
Combine the exponential terms involving $x$ inside the integral:
\begin{equation}
e^K \int_{-\infty}^{\infty} M_\Q(x) \frac{1}{\sqrt{2\pi\tau}} \exp\left(Gx -\frac{x^2}{2\tau}\right) \ud x = e^{r\tau}
\end{equation}
We complete the square for the term $Gx - \frac{x^2}{2\tau}$ in the exponent:
\begin{align}
Gx - \frac{x^2}{2\tau} &= -\frac{1}{2\tau} (x^2 - 2G\tau x) \\
&= -\frac{1}{2\tau} (x^2 - 2G\tau x + (G\tau)^2 - (G\tau)^2) \quad \text{(add and subtract } (G\tau)^2 \text{)} \\
&= -\frac{1}{2\tau} \left( (x - G\tau)^2 - (G\tau)^2 \right) \\
&= -\frac{(x - G\tau)^2}{2\tau} + \frac{(G\tau)^2}{2\tau} = -\frac{(x - G\tau)^2}{2\tau} + \frac{G^2\tau}{2}
\end{align}
Substitute this back into the integral:
\begin{equation}
e^K \int_{-\infty}^{\infty} M_\Q(x) \frac{1}{\sqrt{2\pi\tau}} \exp\left( -\frac{(x - G\tau)^2}{2\tau} + \frac{G^2\tau}{2} \right) \ud x = e^{r\tau}
\end{equation}
The term $\exp(G^2\tau/2)$ does not depend on $x$ and can be factored out of the integral:
\begin{equation}
e^K \exp\left(\frac{G^2\tau}{2}\right) \int_{-\infty}^{\infty} M_\Q(x) \frac{1}{\sqrt{2\pi\tau}} \exp\left( -\frac{(x - G\tau)^2}{2\tau} \right) \ud x = e^{r\tau}
\end{equation}
This can be written as:
\begin{equation} \label{eq:expect_P_star_form}
e^{K + G^2\tau/2} \E_{P^*}[M_\Q(X)] = e^{r\tau}
\end{equation}
where $\E_{P^*}[\cdot]$ denotes the expectation with respect to a random variable $X$ that follows a Normal distribution with mean $G\tau$ and variance $\tau$, i.e., $X \sim \Normal(G\tau, \tau)$.
Now, take the natural logarithm of both sides of \cref{eq:expect_P_star_form}:
\begin{equation}
K + \frac{G^2\tau}{2} + \ln \left( \E_{P^*}[M_\Q(X)] \right) = r\tau
\end{equation}
Substitute back the expressions for $K = (\mu - \tfrac{1}{2}\sigma^2 - \tfrac{1}{2}\gamma_D^2 \sigma^2)\tau$ and $G = \sigma(1 - \gamma_D)$:
\begin{equation}
(\mu - \tfrac{1}{2}\sigma^2 - \tfrac{1}{2}\gamma_D^2 \sigma^2)\tau + \frac{(\sigma(1 - \gamma_D))^2\tau}{2} + \ln \left( \E_{P^*}[M_\Q(X)] \right) = r\tau
\end{equation}
Expand the term $\frac{(\sigma(1 - \gamma_D))^2\tau}{2}$:
\begin{equation}
\frac{\sigma^2(1 - \gamma_D)^2\tau}{2} = \frac{\sigma^2(1 - 2\gamma_D + \gamma_D^2)\tau}{2} = \left(\tfrac{1}{2}\sigma^2 - \gamma_D \sigma^2 + \tfrac{1}{2}\gamma_D^2 \sigma^2\right)\tau
\end{equation}
Substitute this into the equation:
\begin{equation}
(\mu - \tfrac{1}{2}\sigma^2 - \tfrac{1}{2}\gamma_D^2 \sigma^2)\tau + (\tfrac{1}{2}\sigma^2 - \gamma_D \sigma^2 + \tfrac{1}{2}\gamma_D^2 \sigma^2)\tau + \ln \left( \E_{P^*}[M_\Q(X)] \right) = r\tau
\end{equation}
Combine the terms involving $\tau$:
\begin{align}
& \mu\tau - \tfrac{1}{2}\sigma^2\tau - \tfrac{1}{2}\gamma_D^2 \sigma^2\tau + \tfrac{1}{2}\sigma^2\tau - \gamma_D \sigma^2\tau + \tfrac{1}{2}\gamma_D^2 \sigma^2\tau + \ln \left( \E_{P^*}[M_\Q(X)] \right) = r\tau \\
& \mu\tau + (-\tfrac{1}{2}\sigma^2\tau + \tfrac{1}{2}\sigma^2\tau) + (-\tfrac{1}{2}\gamma_D^2 \sigma^2\tau + \tfrac{1}{2}\gamma_D^2 \sigma^2\tau) - \gamma_D \sigma^2\tau + \ln \left( \E_{P^*}[M_\Q(X)] \right) = r\tau \\
& \mu\tau + 0 + 0 - \gamma_D \sigma^2\tau + \ln \left( \E_{P^*}[M_\Q(X)] \right) = r\tau \\
& (\mu - \gamma_D \sigma^2)\tau + \ln \left( \E_{P^*}[M_\Q(X)] \right) = r\tau
\end{align}
Now, solve for $\mu$:
\begin{equation}
\mu \tau = r \tau + \gamma_D \sigma^2 \tau - \ln \left( \E_{P^*}[M_\Q(X)] \right)
\end{equation}
Divide by $\tau$ (assuming $\tau > 0$):
\begin{equation}
\mu = r + \gamma_D \sigma^2 - \frac{1}{\tau} \ln \left( \E_{P^*}[M_\Q(X)] \right) \label{eq:mu_derivation_conclusion}
\end{equation}
The expectation $\E_{P^*}[M_\Q(X)]$ is taken with respect to $X \sim \Normal(G\tau, \tau)$, where $G\tau = \sigma(1 - \gamma_D)\tau$.

\begin{theorem}[The No-Arbitrage Condition] \label{thm:no_arbitrage_condition_revised}
For the asset price model defined in \Cref{sec:model_p} to be free of arbitrage, given a set of market prices of risk $\gamma_D$ (for diffusion) and $\eta_{jk}$ (for jumps), the physical drift $\mu$ of the asset $S_t$ must satisfy the following condition:

\begin{equation}
\mu = r + \gamma_D \sigma^2 - \frac{1}{\tau} \ln \left( \E_{N(\sigma(1 - \gamma_D)\tau, \tau)} [M_\Q(X)] \right)
\end{equation}

where $M_\Q(x) = \E_\Q[e^{J^\Q} | \Delta W = x]$ is the conditional Moment Generating Function (MGF) of the jump factor $\exp(J^\Q)$ under the risk-neutral measure $\Q$, evaluated at $\Delta W = x$:

\begin{equation}
M_\Q(x) = \begin{cases} 
q_{1u}(x)e^{\nu_{1u}^\Q + \delta_{1u}^2/2} + q_{1d}(x)e^{\nu_{1d}^\Q + \delta_{1d}^2/2} + q_{10}(x) & \text{if } x < b_d \sqrt{\tau} \\
q_{2u}(x)e^{\nu_{2u}^\Q + \delta_{2u}^2/2} + q_{2d}(x)e^{\nu_{2d}^\Q + \delta_{2d}^2/2} + q_{20}(x) & \text{if } x > b_u \sqrt{\tau} \\
1 & \text{if } b_d \sqrt{\tau} \le x \le b_u \sqrt{\tau} 
\end{cases}
\end{equation}

with the risk-neutral jump probabilities $q_{jk}(x)$ and risk-neutral jump means $\nu_{jk}^\Q$ defined in \cref{prop:jumps_q}. The expectation $\E_{N(\sigma(1 - \gamma_D)\tau, \tau)} [\cdot]$ is taken with respect to a random variable $X$ distributed as $\Normal(\sigma(1 - \gamma_D)\tau, \tau)$, which means:
\begin{equation}
\E_{N(\sigma(1 - \gamma_D)\tau, \tau)} [M_\Q(X)] = \int_{-\infty}^{\infty} M_\Q(x) \frac{1}{\sqrt{2\pi\tau}} \exp\left( -\frac{(x - \sigma(1 - \gamma_D)\tau)^2}{2\tau} \right) \ud x
\end{equation}
\end{theorem}
\begin{proof} 
The detailed derivation provided above, from \cref{eq:main_start_point} to \cref{eq:mu_derivation_conclusion}, constitutes the proof of this theorem. The core of the proof lies in relating the martingale condition under $\Q$ back to an expectation under $\Prob$ via $L_\tau$, and then evaluating this $\Prob$-expectation. The derivation shows that if $\mu$ satisfies the stated condition, then $\E_\Prob[L_\tau S_{t+\tau}/S_t] = e^{r\tau}$, which is equivalent to $\E_\Q[S_{t+\tau}/S_t] = e^{r\tau}$, thus ensuring no arbitrage by \cref{thm:ftap}.
\end{proof}

\begin{remark}[The Guarantee of No Arbitrage]
\Cref{thm:no_arbitrage_condition_revised} provides the explicit "guarantee" against arbitrage. If the parameters of the model under the physical measure $\Prob$ and the chosen market prices of risk are such that the equation for $\mu$ holds, then the constructed measure $\Q$ is an EMM, ensuring by FTAP that the model is arbitrage-free. Different choices of risk premia would lead to different (but still arbitrage-free) physical drifts $\mu$ or, if $\mu$ is fixed, imply certain market prices of risk.
\end{remark}

\section{Discussion} \label{sec:discussion}

The no-arbitrage condition decomposes the required physical drift into three components:
\begin{equation}
\mu = \underbrace{r}_{\text{Risk-free rate}} + \underbrace{\gamma_D \sigma^2}_{\text{Diffusion risk premium}} - \underbrace{\frac{1}{\tau} \ln \left( \E_{N(\sigma(1 - \gamma_D)\tau, \tau)} [M_\Q(X)] \right)}_{\text{Jump risk adjustment}}
\end{equation}

The jump risk adjustment captures the complex impact of state-dependent jumps:
\begin{itemize}
    \item If $\E[M_\Q(X)] > 1$, risk-neutralized jumps have positive expected impact, reducing required drift
    \item If $\E[M_\Q(X)] < 1$, jumps have negative expected impact, requiring higher drift compensation
    \item Different $\eta_{jk}$ allow differentiated pricing of various jump risks
\end{itemize}

This framework enables:
\begin{enumerate}
    \item \textbf{Model Consistency:} Ensuring that any specific parameterization of the model under the physical measure $\Prob$ is internally consistent and does not admit trivial arbitrage strategies.
    \item \textbf{Derivative Pricing:} By establishing the EMM $\Q$, derivative securities can be priced using the principle of risk-neutral valuation, i.e., $\text{Price}_t = \E_\Q[e^{-r(T-t)} \text{Payoff}_T | \F_t]$.
    \item \textbf{Risk Management:} Providing a deeper understanding of how diffusion and state-dependent jump risks interact and how they are priced by the market. This is essential for developing effective hedging strategies and for accurate risk assessment.
\end{enumerate}

The discrete-time approach facilitates direct implementation while building upon established no-arbitrage principles. Extension to continuous time remains an important avenue for future research.

\section{Conclusion} \label{sec:conclusion}

This paper rigorously addresses the theoretical challenge of constructing arbitrage-free models with diffusion-dependent jumps. We have formalized and extended previous work into a comprehensive multi-type jump framework that eliminates arbitrage concerns in one-sided models.

The central achievement is deriving the explicit no-arbitrage condition that precisely links the physical drift to all model parameters and market risk premia. This condition provides a guarantee: any model whose parameters satisfy this relationship is, by construction, arbitrage-free.

Future research directions include investigating the continuous-time limit, developing efficient numerical methods for option pricing and calibration, empirical testing against market data, and extensions to multi-asset scenarios or stochastic volatility models.

\bibliographystyle{amsplain}

\end{document}